\input{aipcheck}

\documentclass[
    ,final            
  ]
  {aipproc}

\layoutstyle{6x9}

\begin{document}

\title{Verification of Z-scaling in pp Collisions at RHIC}


\author{M.Tokarev}{
  address={JINR, 141980 Dubna, Moscow region, Russia}
}

\author{I.Zborovsk\'{y}}{
  ,address={NPI, 25068 \v {R}e\v {z}, Czech Republic}
}


\begin{abstract}
New experimental data on inclusive spectra of identified particles
produced in $pp$ collisions  at the RHIC are used to test $z$-scaling.
Energy and multiplicity independence  of the scaling function
is established.
The RHIC data confirm $z$-scaling
observed  at U70, ISR, SpS and Tevatron energies.
The obtained results are of interest  to search
for new physics phenomena of particle production in high
transverse momentum and high multiplicity region at  RHIC, Tevatron and LHC.

%
%
\end{abstract}

\maketitle


\section{Introduction}

Study of scaling regularities in high energy collisions
 is always
subject of intense experimental and theoretical
investigations \cite{Feynman}-\cite{Brodsky}.
Some scalings can reflect fundamental symmetries in Nature.
Basic principles to study such symmetries  at small scales
are self-similarity, locality and fractality.
New scaling ($z$-scaling) for description of high-$p_T$ particle
production in inclusive reactions was established in \cite{Z}.
Properties of $z$-presentation of numerous experimental data
confirm self-similarity, locality  and fractality of hadron interactions
at high energies.
The Relativistic Heavy Ion Collider (RHIC) at the
Brookhaven National Laboratory (BNL) gives wide possibilities to
perform experimental measurements and test scaling regularities
in a new physics domain.

We present  results of analysis of new data on high-$p_T$ particle
spectra obtained at the RHIC. The data confirm $z$-scaling
observed at the U70, ISR, SpS and  Tevatron.

\section{z-Scaling}

Search for an adequate, physically meaningful but still sufficiently simple form of the
self-similarity parameter $z$ plays a crucial role in our approach.
For inclusive reactions  we define the scaling variable
\begin{equation}
z = \frac{s^{1/2}_{\bot}}{{\it W }}
\label{eq:r20a}
\end{equation}
as ratio of  the  minimal transverse kinetic energy $
{s^{1/2}_{\bot}}$ of underlying constituent subprocess and
relative number $W$ of such configurations of the colliding system
which can contribute to production of inclusive particle with the
momentum $p$. The number of configurations is expressed via the
multiplicity density $dN/d\eta|_0 $
 at pseudorapidity $\eta=0$ and kinematical characteristics $x_1, x_2$ and $y$
of the subprocess as follows
\begin{equation}
{\it W}=(dN/d\eta|_0)^c\cdot\Omega(x_1,x_2,y),
\label{eq:r20}
\end{equation}
where
\begin{equation}
\Omega(x_1,x_2,y)=(1-x_1)^{\delta_1}(1-x_2)^{\delta_2}(1-y)^{\epsilon}.
\label{eq:r6}
\end{equation}
Here $x_1$ and $x_2$   are momentum fractions of the colliding
objects (hadrons or nuclei). The $y$ is momentum fraction of
outgoing constituent from the subprocess carried by the inclusive
particle. The $\delta_1$, $\delta_2$ and $\epsilon$ are anomalous
fractal dimensions of the incoming and outgoing objects,
respectively. The variable $z$ has character of a  fractal measure
\begin{equation}
z =z_0 \Omega^{-1}.
\label{eq:r5}
\end{equation}
Its divergent part $\Omega^{-1}$ describes  resolution
at which the collision of the constituents can be singled out of inclusive reaction.
With increasing resolution the measure
 $z$ tends to infinity.
The  $x_{1}$, $x_{2}$ and $y$ are determined in a way to minimize
the resolution $\Omega^{-1}(x_{1},x_{2},y)$ taking into account
the energy-momentum conservation of the binary subprocess written
in the form
\begin{equation}
(x_1P_1+x_2P_2-p/y)^2 =(x_1M_1+x_2M_2+m_2/y)^2.
\label{eq:r3}
\end{equation}
Here  $P_1, P_2$ and $M_1, M_2$ are 4-momenta  and masses of the
colliding objects. The $p$ is 4-momentum of the inclusive
particle. The parameter $m_2$ is minimal mass introduced to
satisfy the internal conservation laws (for baryon number,
isospin, strangeness,...).

The relative number $W$ of the configurations which include the
constituent subprocess
 is expressed via entropy of the rest of the colliding system as follows
\begin{equation}
{\it S} = \ln {\it W}.
\label{eq:r20b}
\end{equation}
Using  equations (\ref{eq:r20}) and (\ref{eq:r6}), we get
\begin{equation}
{\it S} = c\ln {\left[dN/d\eta|_0\right]}+
\ln{[(1\!-\!x_1)^{\delta_1}(1\!-\!x_2)^{\delta_2}(1\!-\!y)^{\epsilon}]}
\label{eq:r20d}
\end{equation}
Exploiting   analogy with the thermodynamical formula
\begin{equation}
{\it S} = c_V\ln {T}+ R\ln {V} + const.
\label{eq:r20c}
\end{equation}
we can consider multiplicity density $dN/d\eta|_0$ as a quantity characterizing
"temperature" of the colliding system  and the parameter
$c$ as "heat capacity" of the medium.
The second term in  (\ref{eq:r20d}) is related to volume of the
 configurations
in space of the momentum fractions  which can contribute to
production of the inclusive particle with the momentum $p$. Note
that minimal resolution $\Omega^{-1}$ of the fractal measure $z$
with respect to constituent subprocesses corresponds to maximal
entropy $\it S$ of the rest of the system.

In accordance with self-similarity principle we search for the
scaling function
\begin{equation}
\psi(z) ={1\over{N\sigma_{in}}}{d\sigma\over{dz}}
\label{eq:r4}
\end{equation}
depending on the single variable $z$. Here $\sigma_{in}$ is the inelastic cross
section of the inclusive reaction and $N$ is particle multiplicity.
The  function $\psi(z)$ is expressed in terms of the experimentally
measured inclusive invariant cross section $Ed^3\sigma/dp^3$ and  multiplicity density
$dN/d\eta$ as follows
\begin{equation}
\psi(z) = -{ { \pi s} \over { (dN/d\eta) \sigma_{in}} } J^{-1} E {
{d^3\sigma} \over {dp^3}  }.
\label{eq:r21}
\end{equation}
Here $s$ is the center-of-mass collision energy squared. The
Jacobian J of transformation to the variables $(z,\eta)$ depends
on the momentum $p$ of the inclusive particle. The $\psi(z)$ has
meaning of probability density to produce inclusive particle with
the corresponding value of the variable $z$.

\section{Z-scaling at RHIC}

We analyze experimental data on minimum bias $pp$ spectra of
different hadrons $(h^{\pm}, \pi^0, \pi^-, K_S^0)$ measured at the RHIC.
Comparison of the RHIC  data with data obtained at the U70, ISR, SpS and  Tevatron
is used to test $z$-scaling.

\begin{figure}
  \includegraphics[height=.2\textheight]{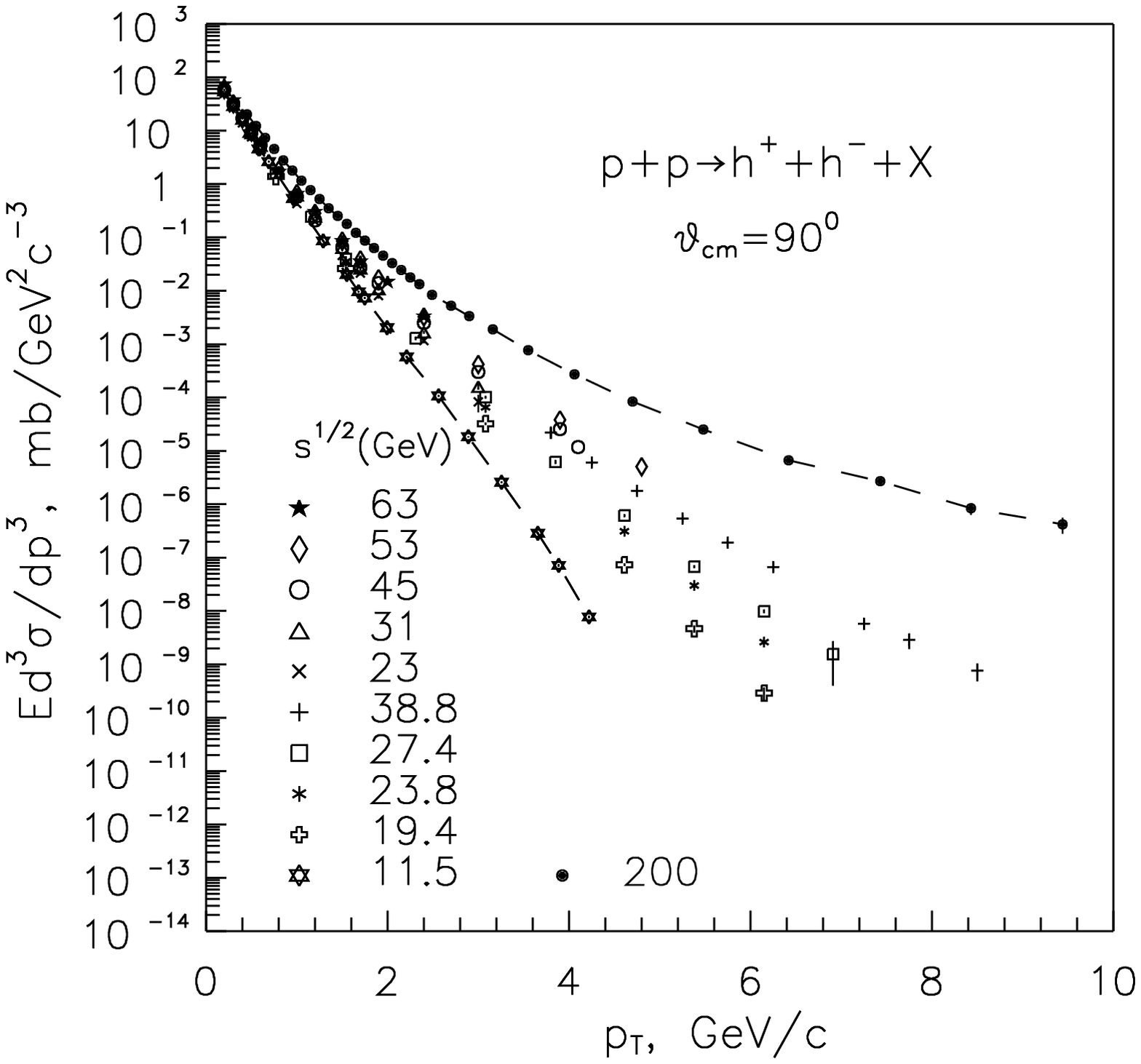}
\hspace*{15mm}  \includegraphics[height=.2\textheight]{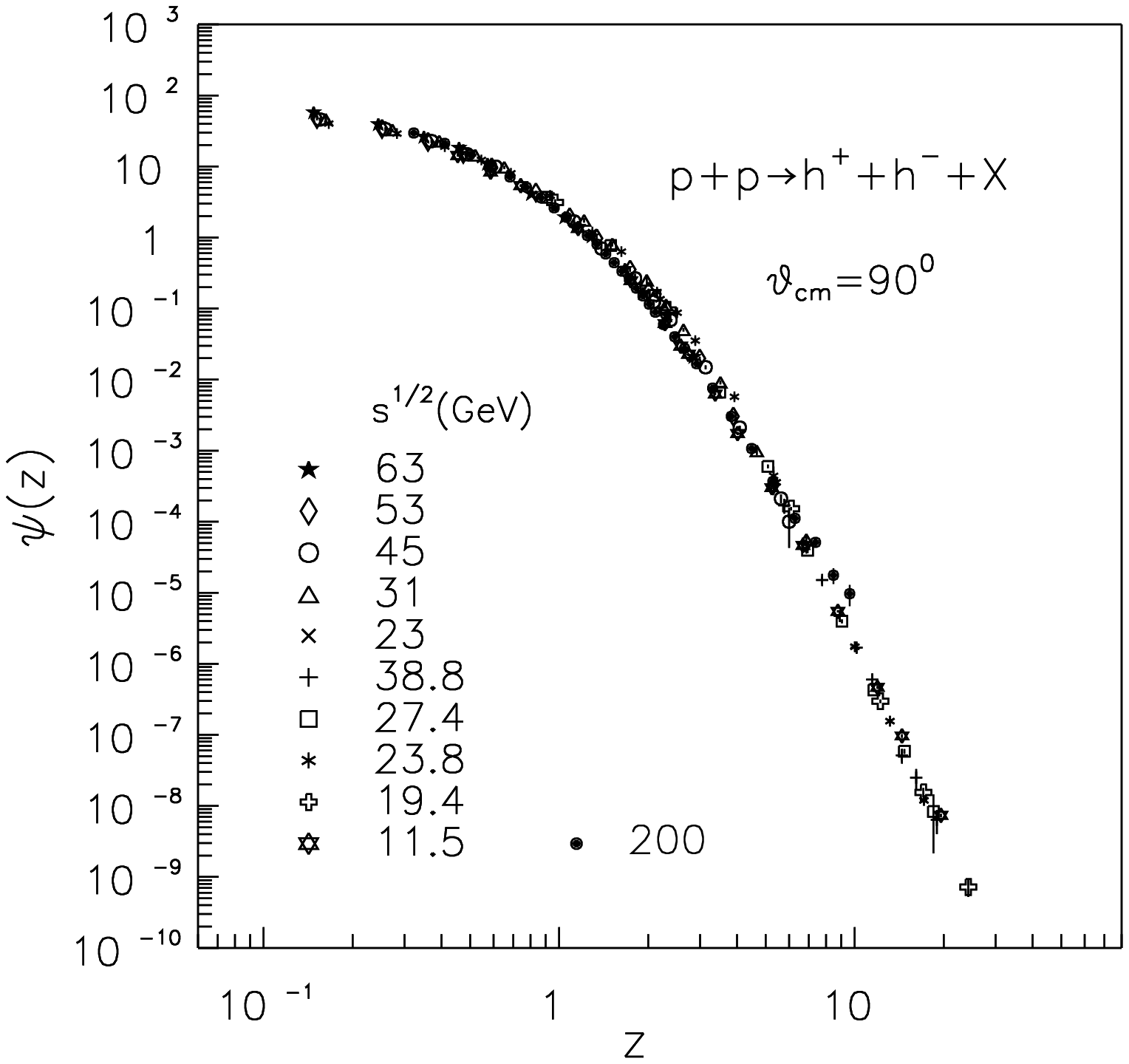}
  \caption{
(a) Inclusive  cross sections
  of charged hadrons produced in $pp$ collisions
at $\sqrt s = 11.5-63$ and 200~GeV and  $\theta_{cm} \simeq 90^{0}$
as a functions of the transverse momentum $p_T$.
The experimental data  are taken from
\cite{Protvino}-\cite{Alper}
and \cite{Adams}.
(b) The corresponding scaling function.
}
\end{figure}

\begin{figure}
  \includegraphics[height=.2\textheight]{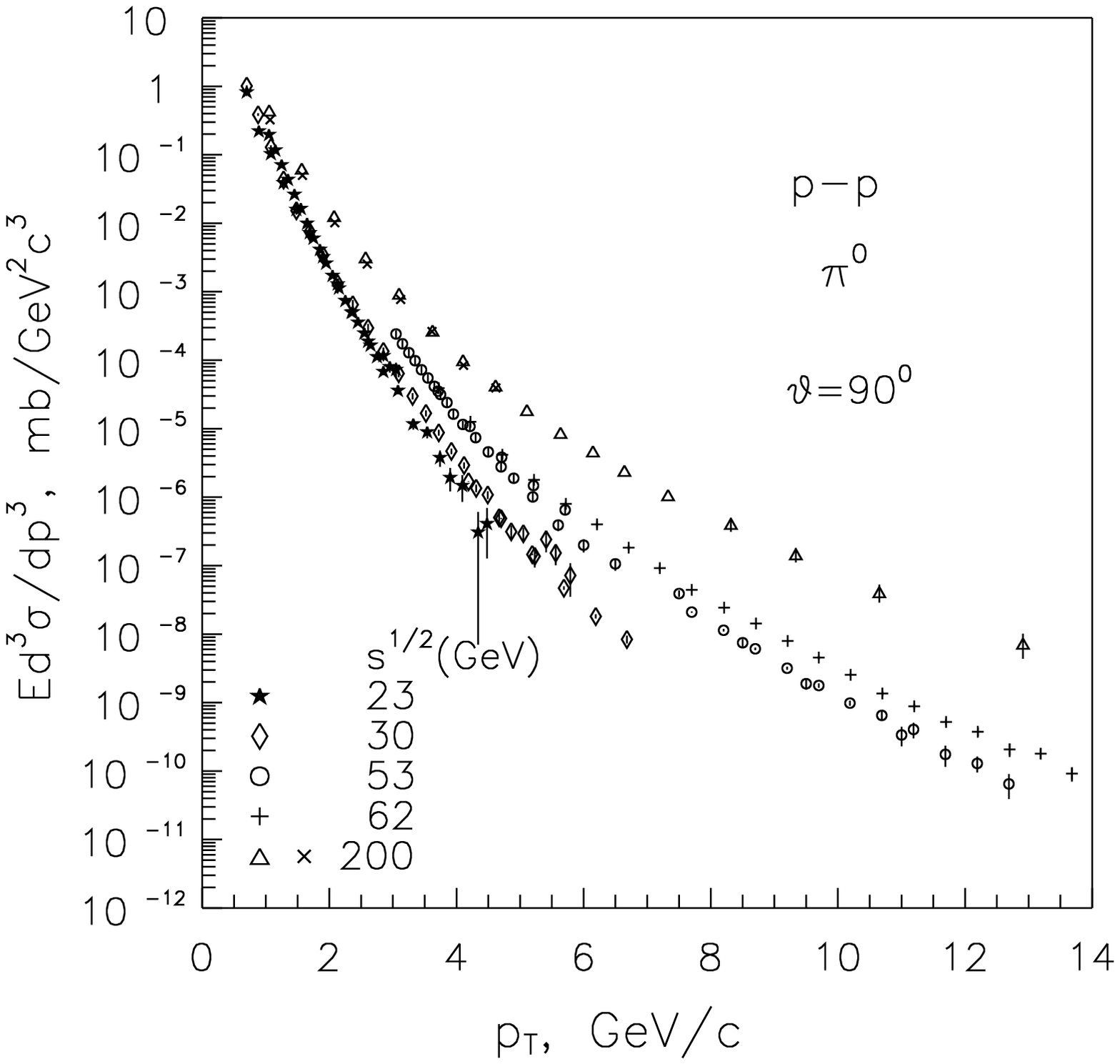}
\hspace*{15mm}  \includegraphics[height=.2\textheight]{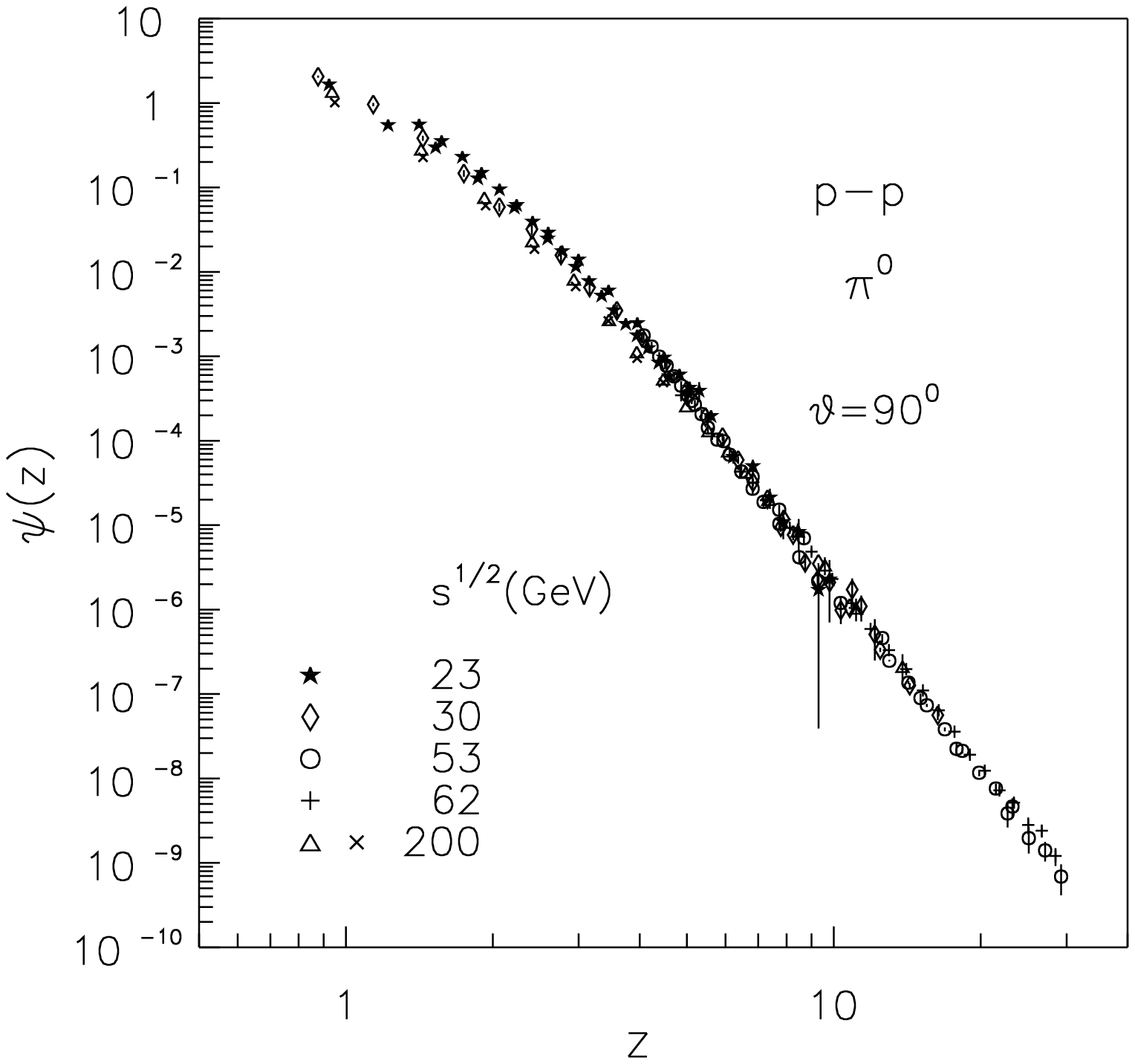}
  \caption{
The dependence of  the inclusive cross section of $\pi^0$-meson production
on the transverse
momentum $p_{T}$ in $pp$ collisions at $\sqrt s = 30,53,62$ and 200~GeV
and the angle $\theta_{cm}$ of $90^0$.
The experimental data  are taken from
\cite{Angel,Kou1,Kou3,Lloyd,Eggert} and \cite{Phenix}.
(b) The corresponding scaling function.
}
\end{figure}

\subsection{Energy independence of $\psi(z)$}

 The high-$p_T$ spectra of charged hadrons  produced in $pp$
collisions at the energy $\sqrt s = 200$~GeV within $|\eta|<0.5$
were measured by the STAR Collaboration \cite{Adams}.
The  inclusive cross sections  obtained
at the U70 \cite{Protvino}, Tevatron \cite{Cronin,Jaffe},  ISR \cite{Alper} and RHIC
are presented in Fig.1a.
The spectra have  strong energy dependence at high $p_T$.
Fig.1b shows $z$-presentation of the same data.
The scaling function   demonstrates  energy independence
and power law, $\psi(z)\sim z^{-\beta}$, for $z>4$.
Results of analysis of the STAR data at $\sqrt s = 200$~GeV
confirm  $z$-scaling observed at lower energies.

\begin{figure}
  \includegraphics[height=.2\textheight]{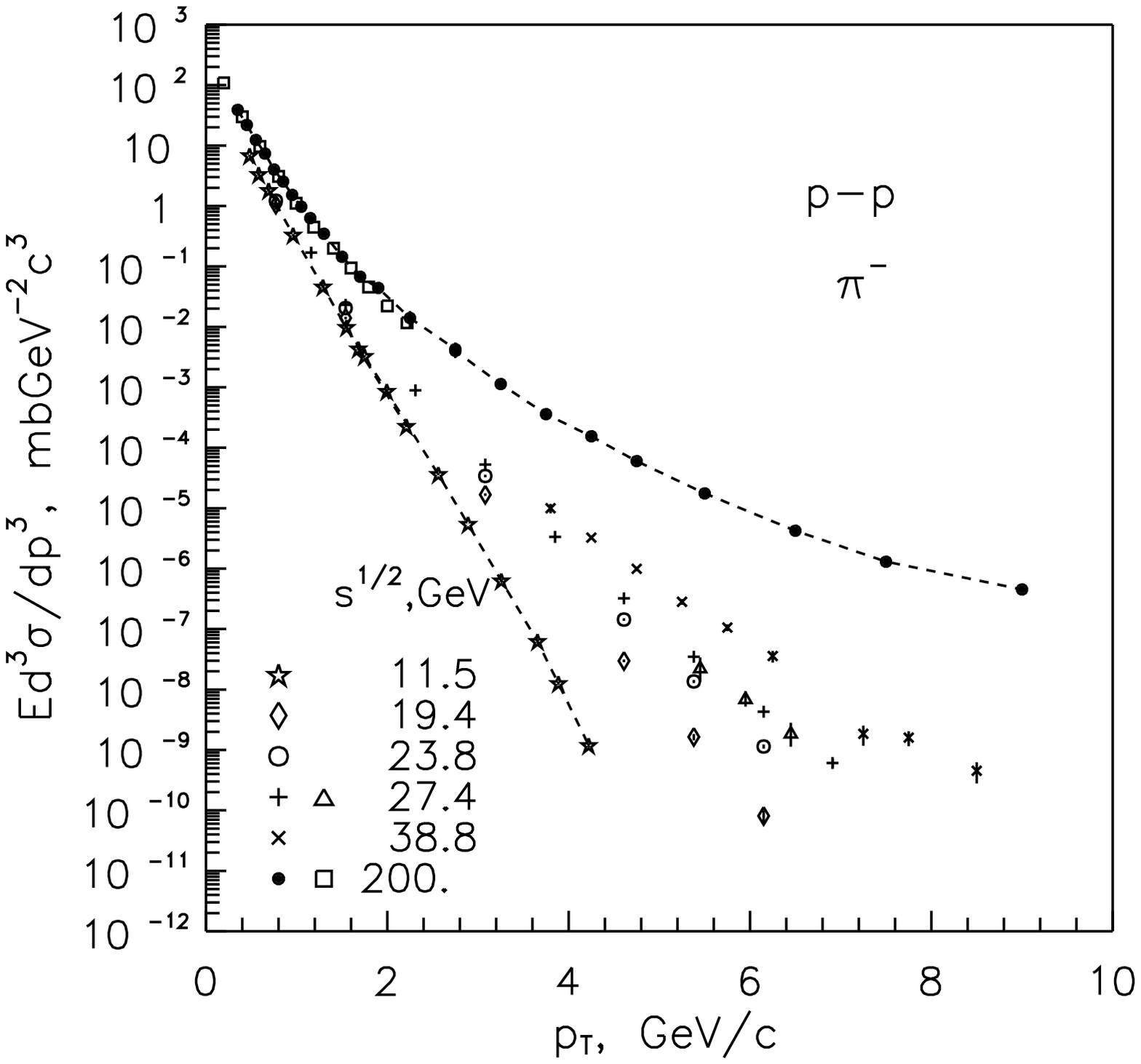}
\hspace*{15mm}  \includegraphics[height=.2\textheight]{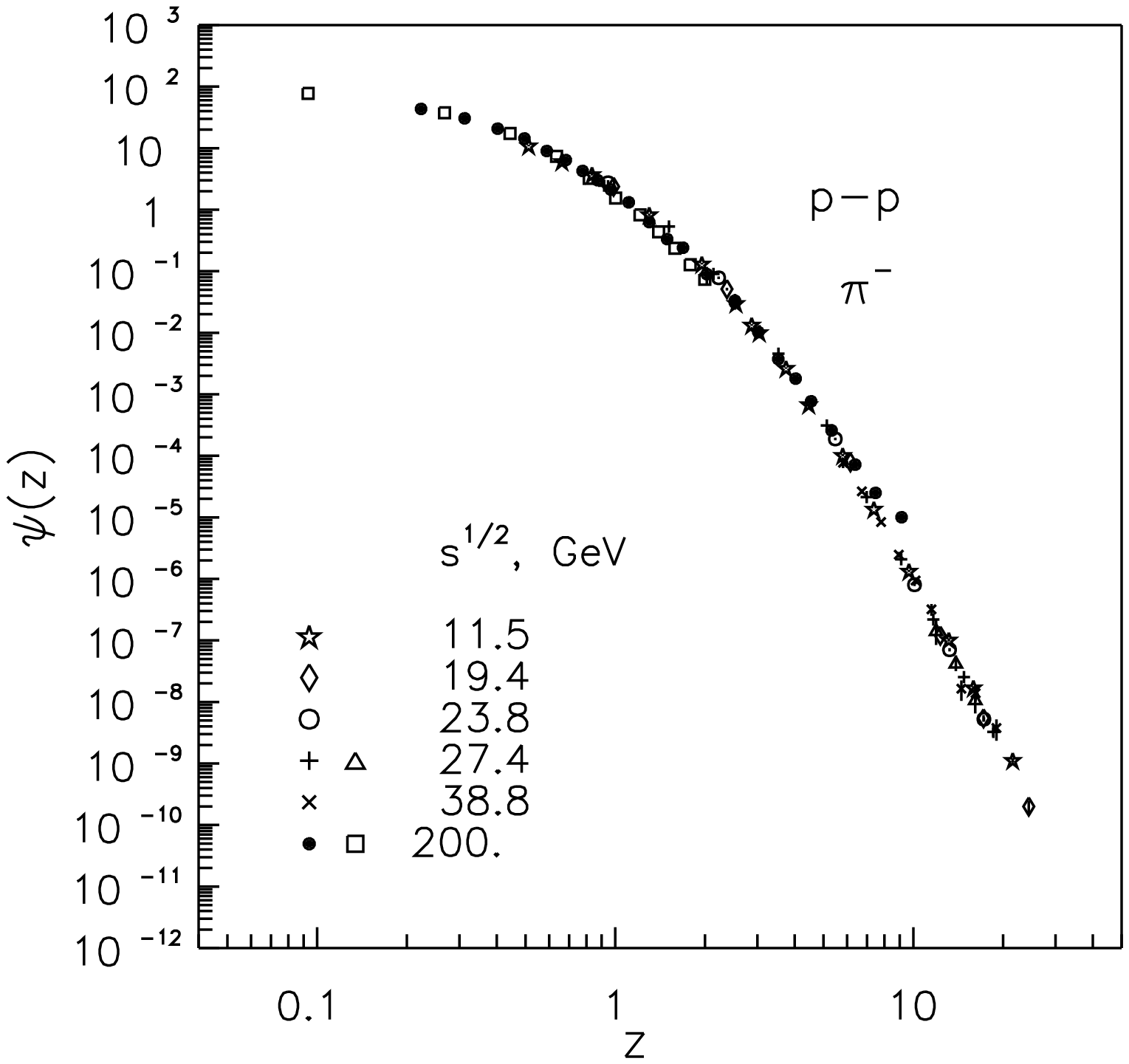}
  \caption{(a) Inclusive  cross sections
  of charged hadrons produced in $pp$ collisions
at $\sqrt s = 11.5-38.8$ and 200~GeV and  $\theta_{cm} \simeq 90^{0}$
as a functions of the transverse momentum $p_T$.
The experimental data  are taken from
\cite{Protvino,Cronin,Jaffe} and \cite{Adams}.
(b) The corresponding scaling function.
}
\end{figure}

\begin{figure}
  \includegraphics[height=.2\textheight]{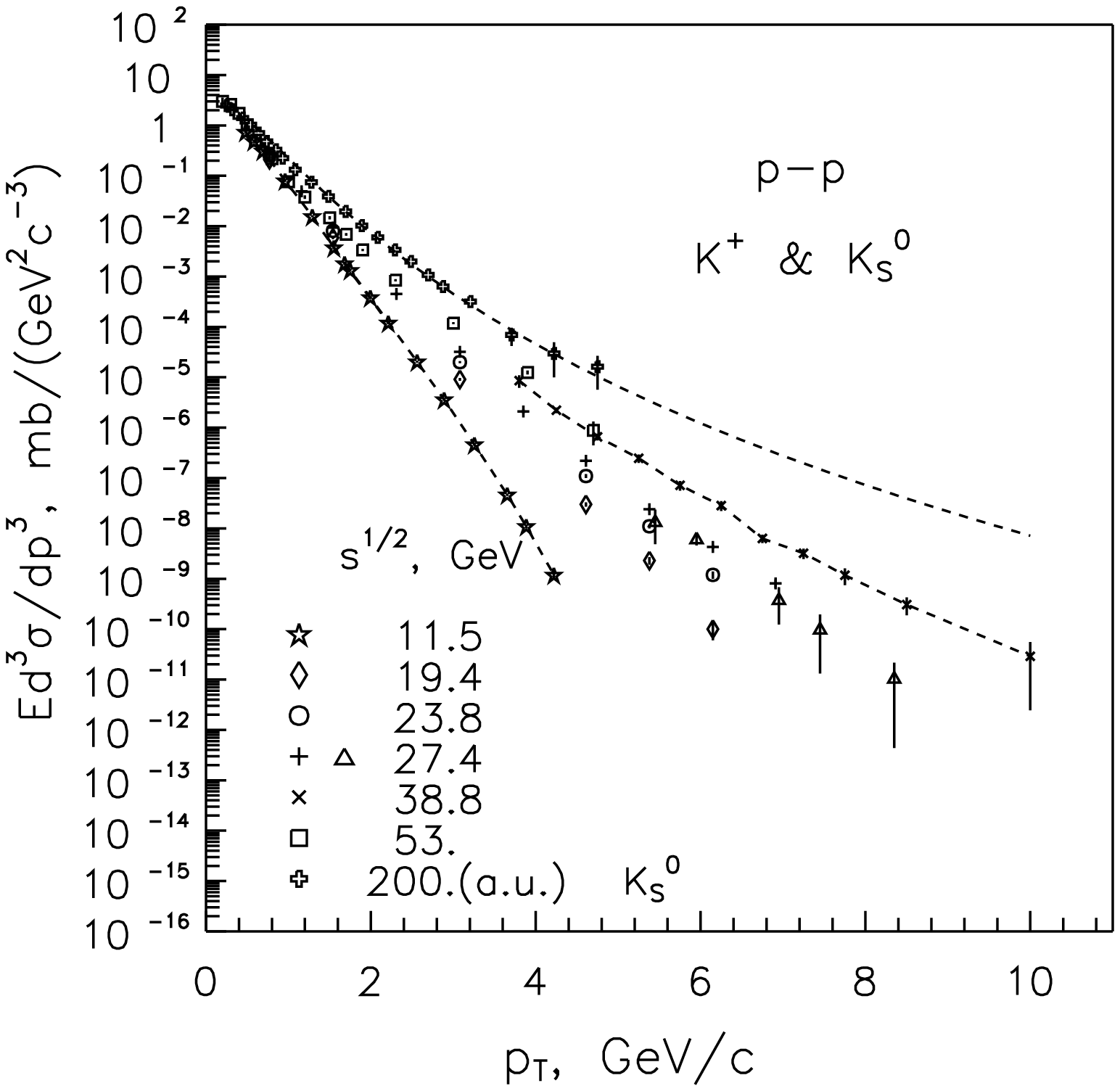}
\hspace*{15mm}  \includegraphics[height=.2\textheight]{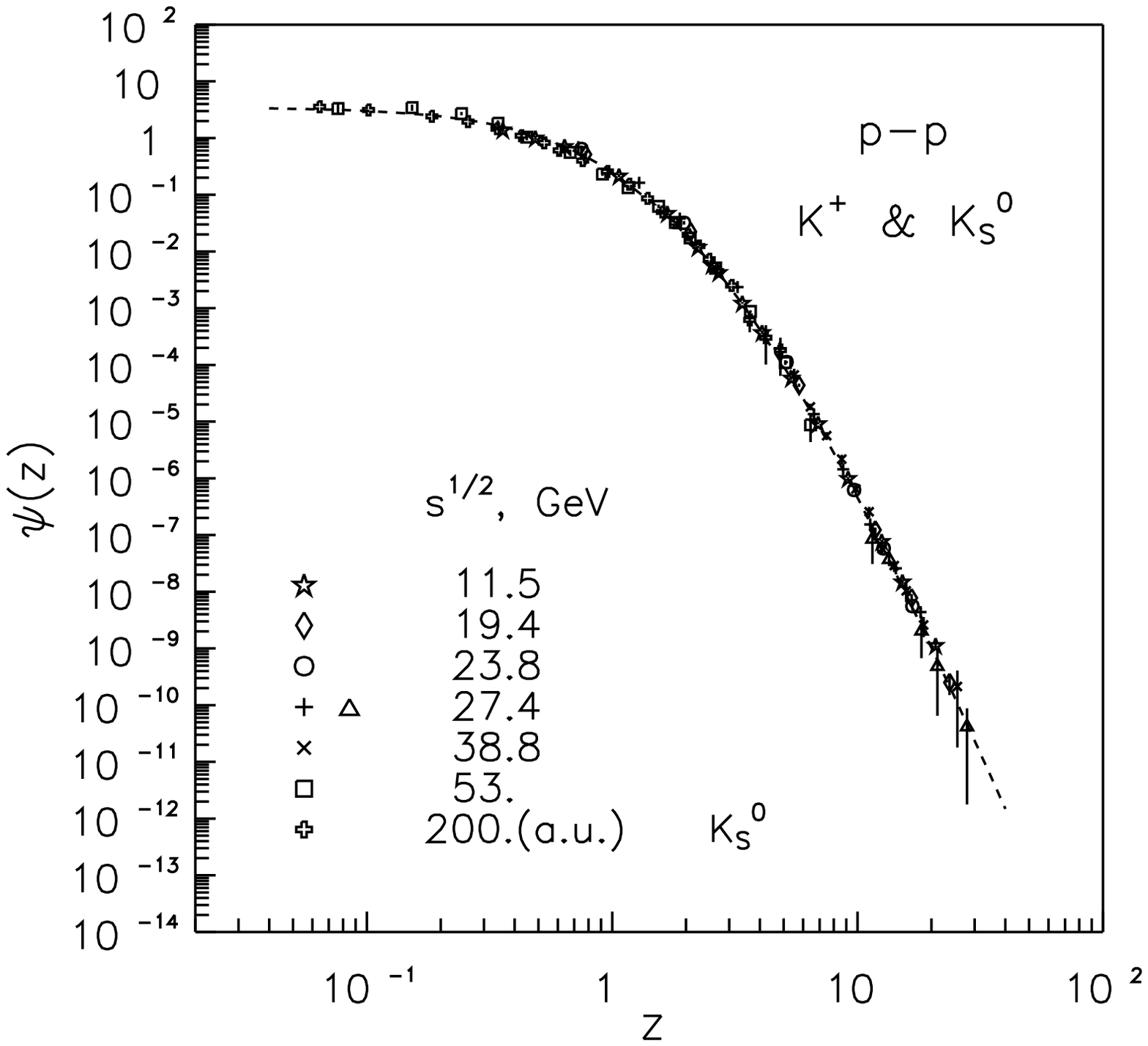}
  \caption{
(a) The inclusive  cross sections of  $K^+$- and  $K_{S}^0$-mesons
produced  in $pp$ collisions in the central rapidity range
as a function of the transverse momentum at
$\sqrt s = 11.5-53$~GeV  and 200~GeV. Experimental data are taken
from \cite{Protvino,Cronin,Jaffe,Alper} and \cite{Heinz}.
(b) The corresponding scaling function.
  }
\end{figure}

The PHENIX Collaboration measured the inclusive spectrum
of $\pi^0$-mesons produced in $pp$ collisions
 at  $\sqrt s = 200$~GeV for $|\eta|<0.35$ and  $p_T$ up to 13~GeV/c
 \cite{Phenix}.
 The $p_T$- and $z$-presentations of data for $\pi^0$-meson spectra obtained at ISR
\cite{Angel,Kou1,Kou3,Lloyd,Eggert} and RHIC  are shown in Figs.2a and 2b, respectively.
Energy dependence of the inclusive cross sections increases with $p_T$.
The PHENIX data at $\sqrt s=200$~GeV
 confirm energy independence  and power law of the scaling function
 for $\pi^0$-mesons.

The STAR Collaboration measured the inclusive spectrum
of $\pi^-$-mesons produced in $pp$ collisions
 at $\sqrt s = 200$~GeV for $|\eta|<0.5$  up to $p_T=9$~GeV/c
\cite{Barannikova}. The data obtained at the STAR,
 U70 \cite{Protvino} and  Tevatron \cite{Cronin,Jaffe} are shown in Fig.3a.
The energy dependence of the $p_T$-spectra is in contrast with the scaling
depicted  in Fig.3b.

The $p_T$-spectra and the scaling function $\psi(z)$
 for data on $K^+$ and $K_S^0$-mesons obtained
at the U70 \cite{Protvino}, Tevatron \cite{Cronin,Jaffe}, ISR \cite{Alper}
and RHIC \cite{Heinz} are presented in Figs. 4a
and 4b, respectively. The shape of the scaling function
for $K_S^0$ is found to be in good agreement with $\psi(z)$ for $K^+$-mesons.
Using  parametrization of $\psi(z)$,
 the dependence of inclusive spectrum of $K_S^0$-mesons on transverse momentum
at $\sqrt s = 200$~GeV is plotted by the dashed line in Fig.4a.

\begin{figure}
  \includegraphics[height=.2\textheight]{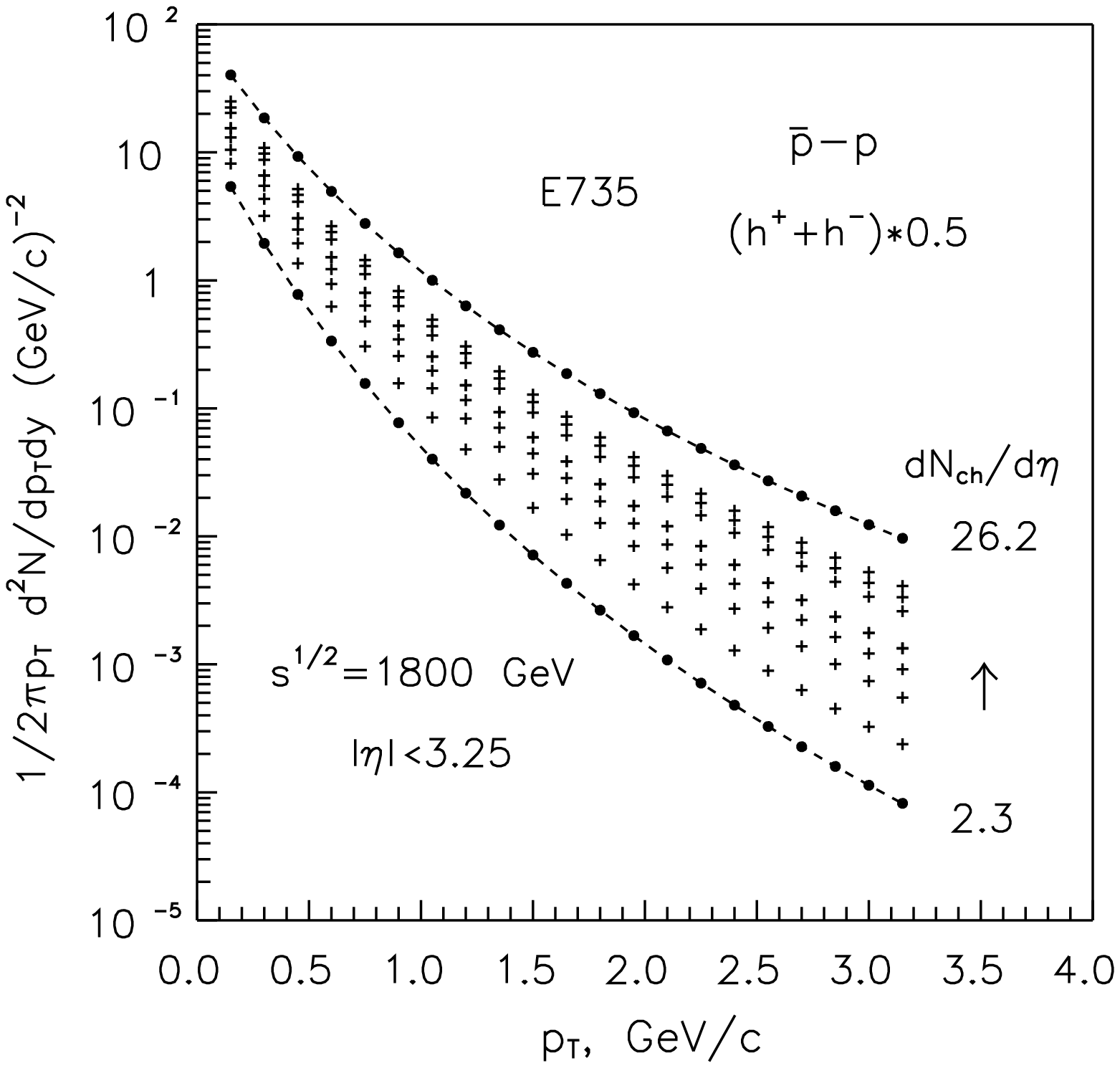}
\hspace*{15mm}  \includegraphics[height=.2\textheight]{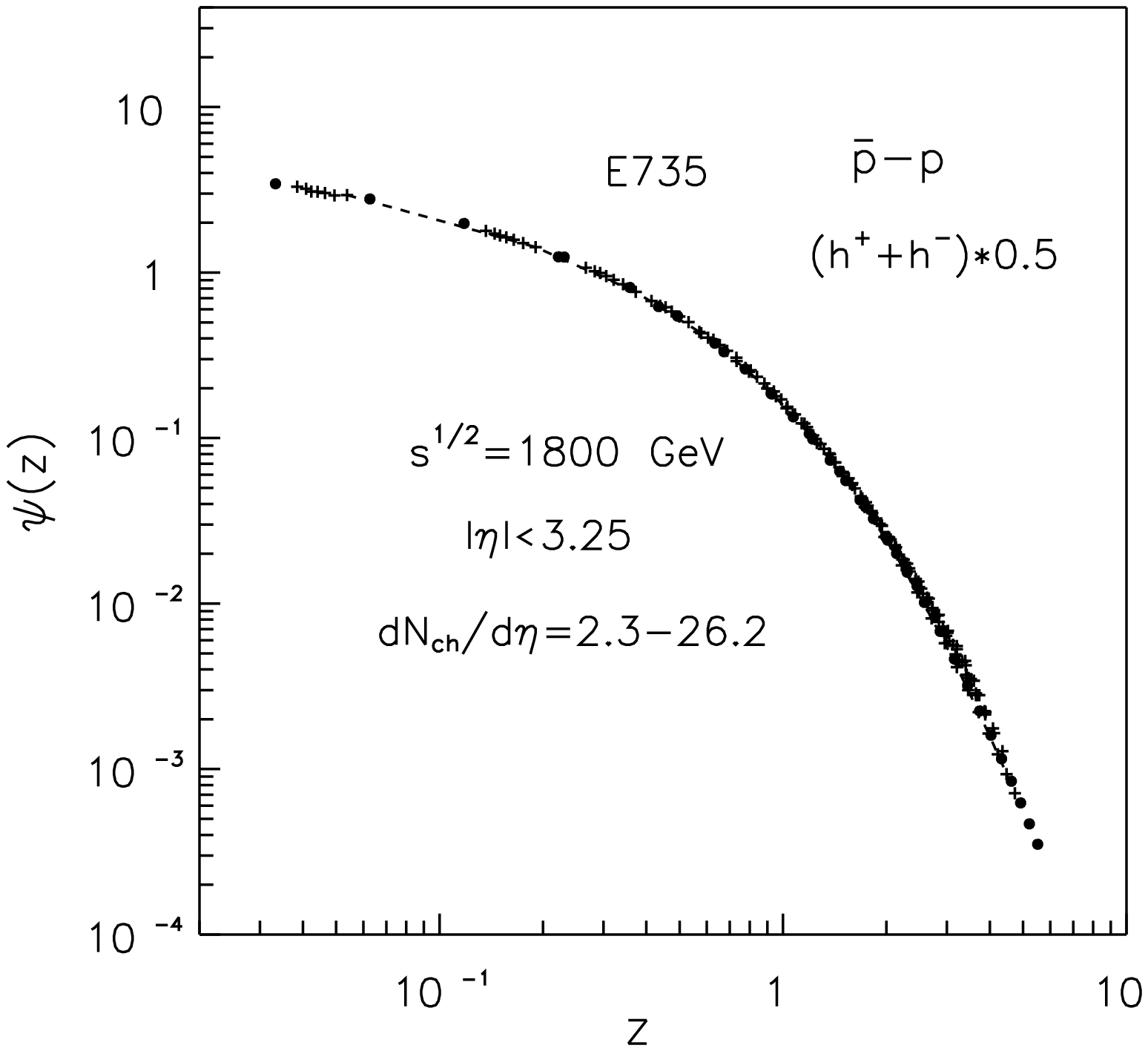}
  \caption{
a) Multiplicity dependence of charged hadron spectra
  in $\bar pp$ collisions  at $\sqrt s=1800$~GeV.
 Experimental data are obtained by the E735 Collaboration \cite{E735}.
(b) The corresponding scaling function.
}
\end{figure}

\begin{figure}
  \includegraphics[height=.2\textheight]{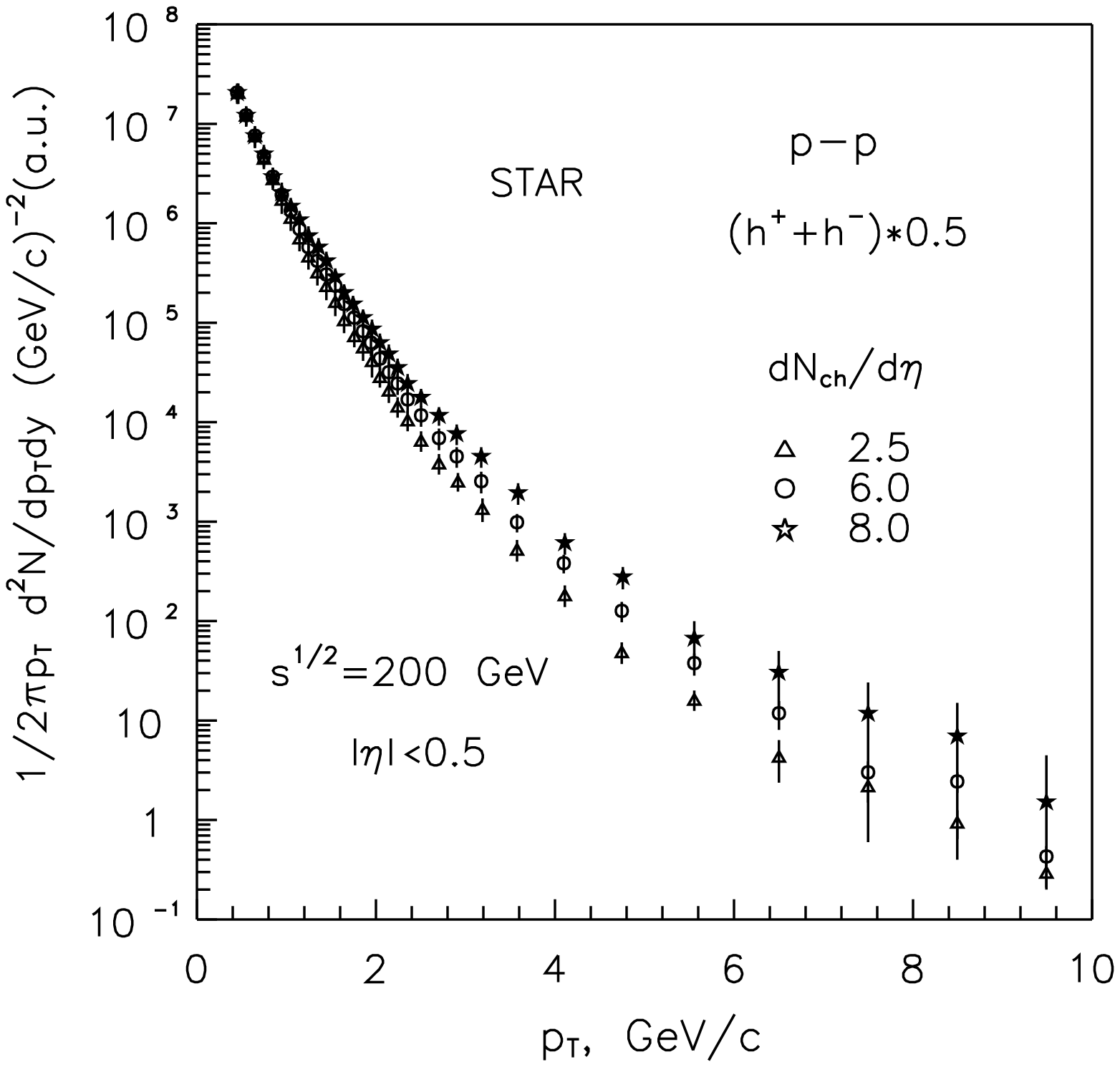}
\hspace*{15mm}  \includegraphics[height=.2\textheight]{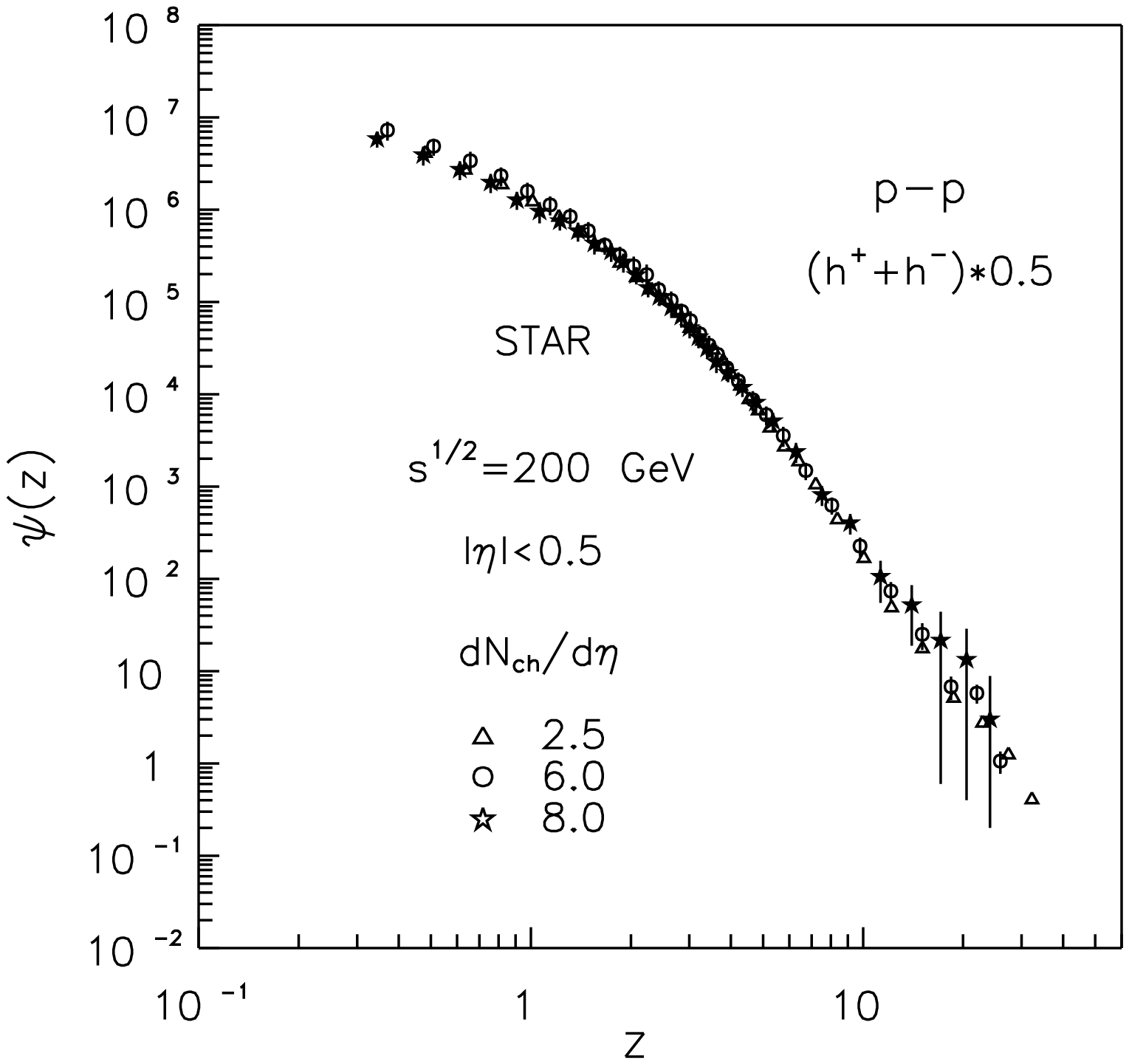}
  \caption{
(a) Multiplicity dependence of charged hadron spectra
  in $ pp$ collisions  at $\sqrt s=200$~GeV.
 Experimental data are obtained by the STAR Collaboration \cite{STAR}.
(b) The corresponding scaling function.
}
\end{figure}

\vskip -0.5cm

\subsection{Multiplicity independence of $\psi(z)$}

We analyze data on charge hadron production
in $pp$ and  $p\bar{p}$  collisions
at different multiplicities and energies.
The E735 Collaboration  measured
the multiplicity dependence of charged hadron spectra
in  $p\bar p$ collisions at  $\sqrt{s}=1800~$GeV
for $dN_{ch}/d\eta=2.3-26.2$, $|\eta|<3.25$ and $p_T = 0.15-3~$GeV/c \cite{E735}.
Strong dependence of the spectra on multiplicity is shown in Fig.5a.
The $z$-presentation of the data is plotted in Fig.5b.
Independence of the scaling function $\psi(z)$ on multiplicity
was found for same value of $c=0.25$.

The STAR Collaboration measured multiplicity dependence of  inclusive spectra
of charged hadrons produced in $pp$ collisions at $\sqrt s = 200$~GeV
for $|\eta|<0.5$ \cite{STAR}.
Fig.6a demonstrates  strong dependence of the spectra on multiplicity.
The STAR data confirm  multiplicity independence
of the scaling function  $\psi(z)$ established in $p\bar{p}$ collisions at higher energies.
The value of the heat capacity $c=0.25$ is found to be the same as for UA1\cite{UA1},
E735 and CDF \cite{CDF_I} data.

Additional confirmation of the $z$-scaling was obtained at RHIC.
The scaling manifests
self-similarity and fractality  in hadron interactions
at high energies.



\begin{theacknowledgments}
 The authors would like to
thank  Yu.Panebratsev for his support of this work.
The investigations have been partially supported by the IRP
AVOZ10480505 and by the Grant Agency of the Czech Republic under
the contract No. 202/04/0793.
\end{theacknowledgments}



\bibliographystyle{aipproc}   

\bibliography{sample}

\IfFileExists{\jobname.bbl}{}
 {\typeout{}
  \typeout{******************************************}
  \typeout{** Please run "bibtex \jobname" to optain}
  \typeout{** the bibliography and then re-run LaTeX}
  \typeout{** twice to fix the references!}
  \typeout{******************************************}
  \typeout{}
 }



\end{document}